%% file: main.tex
\documentclass[acmtog]{acmart}

\usepackage{subcaption}
\usepackage{graphicx}
\usepackage{enumitem}
\usepackage{xcolor}
\usepackage{multirow}
\usepackage{xcolor,colortbl}

\usepackage{amsmath, amssymb, amsfonts}
\usepackage[ruled,vlined,linesnumbered]{algorithm2e}

\newif\ifshowcomments
\showcommentstrue

\definecolor{szcolor}{rgb}{0.2,0.7,0.0}

\AtBeginDocument{%
  }

\copyrightyear{2026}
\acmYear{2026}
\setcopyright{cc}
\setcctype{by}
\acmConference[SA Conference Papers '26]{SIGGRAPH Asia 2026 Conference Papers}{December 01--04, 2026}{Kuala Lumpur, Malaysia}
\acmBooktitle{SIGGRAPH Asia 2026 Conference Papers (SA Conference Papers '26), December 01--04, 2026, Kuala Lumpur, Malaysia}
\acmDOI{10.1145/3829340.3842354}
\acmISBN{979-8-4007-2842-6/2026/12}

\acmSubmissionID{2451}

\begin{document}

\title{Providing Rapid Design Feedback for 3D Obstacle Course Games \\ Using Constrained Solvability Queries}

\author{Zander Majercik}
\email{majercik@stanford.edu}
\orcid{}
\affiliation{%
  \institution{Stanford University}
  \city{Stanford}
  \state{California}
  \country{USA}
}
\affiliation{%
  \institution{Roblox}
  \city{San Mateo}
  \state{California}
  \country{USA}
}
\author{Sharon Zhang}
\email{szhang25@stanford.edu}
\orcid{0000-0002-6738-8906}
\affiliation{%
  \institution{Stanford University}
  \city{Stanford}
  \state{California}
  \country{USA}
}
\affiliation{%
  \institution{Roblox}
  \city{San Mateo}
  \state{California}
  \country{USA}
}

\author{William Wang}
\email{}
\orcid{}
\affiliation{%
  \institution{Stanford University}
  \city{Stanford}
  \state{California}
  \country{USA}
}

\author{Tejan Karmali}
\email{}
\orcid{}
\affiliation{%
  \institution{Stanford University}
  \city{Stanford}
  \state{California}
  \country{USA}
}

\author{Fangjun Zhou}
\email{}
\orcid{}
\affiliation{%
  \institution{Stanford University}
  \city{Stanford}
  \state{California}
  \country{USA}
}

\author{Yucheng Yuan}
\email{}
\orcid{}
\affiliation{%
  \institution{Stanford University}
  \city{Stanford}
  \state{California}
  \country{USA}
}

\author{Jean-Peïc Chou}
\email{}
\orcid{}
\affiliation{%
  \institution{Stanford University}
  \city{Stanford}
  \state{California}
  \country{USA}
}

\author{Maneesh Agrawala}
\email{}
\orcid{}
\affiliation{%
  \institution{Stanford University}
  \city{Stanford}
  \state{California}
  \country{USA}
}
\affiliation{%
  \institution{Roblox}
  \city{San Mateo}
  \state{California}
  \country{USA}
}

\author{Kayvon Fatahalian}
\email{}
\orcid{}
\affiliation{%
  \institution{Stanford University}
  \city{Stanford}
  \state{California}
  \country{USA}
}
\affiliation{%
  \institution{Roblox}
  \city{San Mateo}
  \state{California}
  \country{USA}
}

\renewcommand{\shortauthors}{Majercik et al.}

\begin{abstract}

\input{abstract}

\end{abstract}



\begin{teaserfigure}
    \centering
    \includegraphics[width=\textwidth]{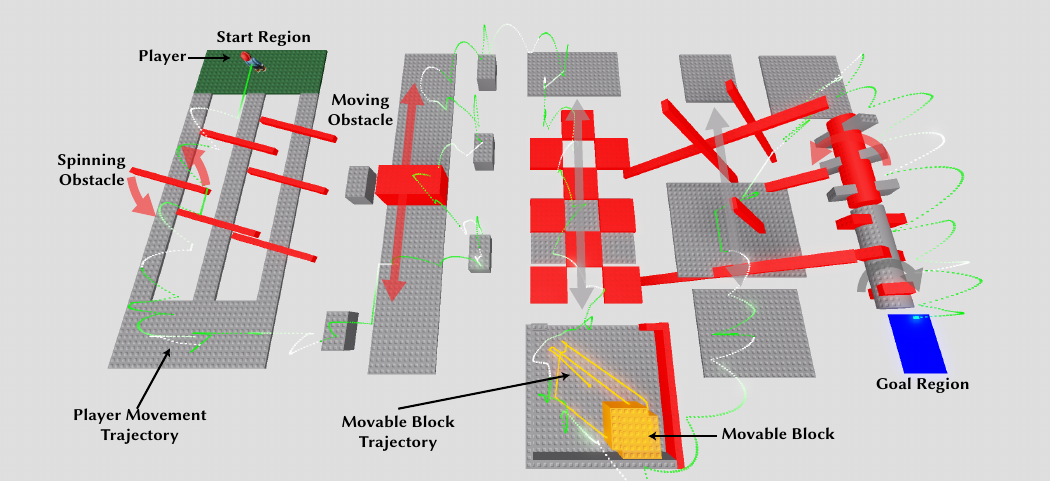}
    \vspace{-2em}
    \caption{Our system provides designers with visualizations of possible play outcomes as they edit a 3D obstacle course video game level. Starting regions are shown in green, goal regions shown in blue, and surfaces that cause player damage are red. Parts of the obstacle may move or rotate to create additional timing challenges for the player. In this example, the player must traverse a series of platforms while dodging deadly red obstacles, and even push a yellow block up to a tall wall in order to jump over it. A solution trajectory automatically found by our system is visualized in green and white. Designers receive this feedback quickly after each edit, allowing them to rapidly iterate towards a desired user experience. A full playthrough of this obstacle is available in a supplemental video.}
    \label{fig:teaser}
\end{teaserfigure}

\maketitle

\input{introduction.tex}

\input{related.tex}

\input{queries.tex}

\input{method.tex}

\input{results.tex}
\input{discussion.tex}

\input{acknowledgements.tex}

\bibliographystyle{ACM-Reference-Format}
\bibliography{main}

\end{document}

%% file: abstract.tex
We present a system that aids the design of 3D obstacle course games by providing designers with rapid feedback on how obstacles can be solved. Our core contribution is a system for querying for solutions (sequences of player actions) to an obstacle that adhere to designer-specified constraints (e.g., avoid a region, travel through a given waypoint, only perform two jumps, etc.). To solve a wide range of obstacle designs quickly, we author a high-performance implementation of the GoExplore algorithm for exploratory search, and guide search with an obstacle solving agent trained offline using reinforcement learning (RL). To further accelerate search, the system carries out exploration using a custom GPU-accelerated obstacle course game simulator that generates playthrough experience at nearly 14,000$\times$ real time, 60-fps gameplay.  
Through design studies, we demonstrate that the use of constrained solvability queries in a rapid design loop is sufficiently expressive to help designers understand ways an obstacle can be solved or why it cannot be solved. We also show how the tool can let designers answer higher-level questions such as identifying undesirable solution paths and assessing the difficulty of solutions. This allows them to pursue new design directions they did not originally anticipate. Human playtesting of obstacles designed using our system confirms that human players indeed play the obstacles in the manner the designers intended. We release code for our interactive tool, simulator, training setup, and procedural level generation system at \url{https://zandermajercik.github.io/interactive-obstacle-course-feedback/}.

%% file: introduction.tex
\section{Introduction}

It is increasingly possible to generate high-fidelity 3D content: meshes\,\cite{xiang:2024:trellis}, texture maps\,\cite{deng:2024:flashtex}, animations\,\cite{tevet:2023:mdm}, even full scenes\,\cite{GenUSD}, from accessible controls like text prompts. These technologies can substantially lower the cost and skill set required to create richly populated and visually interesting 3D worlds.  However, a designer's ultimate goal is rarely just to produce a high visual fidelity result. Rather, designers intend to create worlds that facilitate a \emph{desired user experience}. In the case of video games, important aspects of an experience relate to how the game is solved. For example, 
is the game too easy or too difficult? Can the player win by doing something undesirable?

To answer such questions, designers rely heavily on human play testing. For example, a designer will play the game as they are creating it to directly experience how their design choices affect gameplay~\cite{sestini:2022:designassistanceIL}. 
However, stopping 
to play a game, even for a few minutes, can be disruptive to a design flow. Using additional play testers to gain further feedback is an even more time-consuming and costly process. To address these challenges, we explore how to enable \emph{interactive design workflows} where the results of automated game playthroughs serve as rapid feedback to the designer about how their design choices may affect play. 

Specifically, we focus on assisting the design of 3D obstacle course games, a popular genre where a player must navigate a series of obstacles while avoiding dangerous surfaces and gaps. In these games, success requires the player to exhibit precise motor skills (press the correct keys at the right time)
and properly reason about game dynamics (e.g., is a certain jump angle sufficient to propel the player over a gap). 
Obstacle course games are popular for both amateur players 
and developers with hundreds of millions of total visits on popular game platforms like Roblox. This makes them a compelling target for new tools that aid the game design experience.

Many questions about the gameplay of obstacle course games pertain to obstacle solvability
(e.g., can an obstacle be solved while avoiding a region, traveling through a sequence of waypoints, or only performing two jumps?).
Therefore, we contribute a system that allows designers to express \emph{solvability queries}: requests to see successful playthroughs of their current design that adhere to specified constraints. To support rapid design loops, we contribute a query execution system that can solve the diverse set of 3D obstacles designers may create (obstacles with any number of blocks, in any configuration, and potentially moving in time) and provide solutions to most queries within seconds. 

To achieve the performance necessary for rapid feedback, we contribute a high-performance parallel implementation of the GoExplore exploratory search algorithm~\cite{Go-Explore:2021}, and optionally guide exploration using an agent trained offline using reinforcement learning (RL).
Our system implements a digital twin architecture: a designer 
creates with a standard game engine design tool
, but receives feedback from agent playthroughs performed in a 
separate GPU-accelerated 
simulator that executes an approximation of the game at nearly a million steps/second. The result is a system that produces agent experience at over 14,000$\times$ real time gameplay at 60 fps, and can find solutions to non-trivial obstacles in seconds.

Through design studies, we demonstrate that 
constrained solvability queries are sufficiently expressive to provide insight on a wide variety of obstacle course design questions. We report on how designers use these queries to solicit playthrough feedback that helps them understand how an obstacle can be solved, why it cannot be solved, and identify undesirable solution paths. 
The responsiveness of the system 
also enabled designers to 
visualize many query results in parallel to 
explore diversity in solution strategies,
reason about difficulty and necessary skills,
and discover new play strategies and design directions they originally did not anticipate.
Finally, human playtesting of the obstacles designed using our system confirms that human players indeed play the obstacles in the manner the designers intended.

%% file: related.tex
\section{Related Work}

\paragraph{Automating game development tasks}
As the capability of AI for games improves,
there is growing interest in using agents to automate game development tasks such as bug finding~\cite{sestini:2022:ccpt,bergdah:2020:testingrl,Zheng:2019:wuji,machado:2018:cicero,balyo:2024:automatinggameregression}, simulate playtesting to estimate difficulty~\cite{Isaksen:2018:survivalAnalysisFlappy,Isaksen:2017:StrategyAndDexterityTetris}, predict measures of fun~\cite{togelius:2007:racing}, or (like our work) assess solvability~\cite{sestini:2022:designassistanceIL, gisslen:2021:adversarialPCG, bauer2013:automated,reachability}.  Our work is inspired by prior investigations of providing automated game solvability feedback, but we needed to innovate to provide a robust, high-performance solvability solution for 3D obstacle course games, a space of games that far exceeds the diversity and complexity of games attempted in prior work.

\paragraph{Navigation and pathfinding solutions.} 3D obstacle course games present the challenges of a navigating a continuous game state space with complex
dynamic scene geometry.
Many existing approaches train policies to navigate obstacles based on scene observations~\cite{levine:2011:space-time,gisslen:2021:adversarialPCG}.
To our knowledge, no open large-scale training dataset of 3D obstacle games currently exists, and our own attempts to train policies on procedurally generated data failed to reliably generalize to novel human-designed obstacle course games (see supplement Sec. 4.3).

Another approach to navigation problems is to extract a discrete graph representation of the game state space (e.g., a navigation mesh) and apply search-based techniques like A* \cite{a-star} and variants\cite{D-star-lite,LPA-star}. Unfortunately, creating an accurate navigation mesh for a complex 3D obstacle course with moving parts and physics-based dynamics is an overly complex problem that is not viable in our interactive editing setting where a designer is constantly modifying the obstacle and seeks immediate feedback.

In addition to A* search, previous works have used tree search methods such as RRT and MCTS to search for gameplay trajectories within a 2D platformer level. Suetake et. al design a gameplay testing tool that visualizes trajectories (similar to ours) using variants of MCTS to represent different gameplay strategies that is effective on 2D platformer games~\cite{suetake-2020}. Tremblay et al. provide analysis of multiple tree search algorithms (A*, RRT, and MCTS) for finding solution trajectories in 2D platformer games~\cite{tremblay2014jump}. Their analysis finds a tradeoff between solution speed and solution diversity over the three tested algorithms, with the fastest technique solving most levels in seconds. However, while the discretization of 2D space on which these methods depend is similar to ours, we required significant modification is required to adapt to the complexity of 3D obstacle course games.

\paragraph{Interactive design feedback.} Interactive design feedback enables designers to quickly explore design spaces subject to specific constraints and use the information to make edits that iterate towards their design goals. The Design Gallery interface computes and visualizes a subset of visual outputs to help users find specific parameter settings to achieve a desired output~\cite{designgalleries}. Many worlds-browsing allows users to filter large spaces of physical simulations using positive and negative spatiotemporal queries~\cite{manyworlds2007, manyworlds2022}. Prior work in sketch-based design leverage user-drawn strokes to surface possible 2D images and 3D forms~\cite{shadowdraw,blockAndDetail,juxtaform}. 
Designing an obstacle involves defining obstacle appearance and the space of possible gameplay. Prior 2D platformer design tools use game rhythm~\cite{launchpad1,launchpad2} and reachability analysis~\cite{reachability} to help game designers evaluate whether their game play design goals are met, though these systems do not extend to 3D games.

%% file: queries.tex
\section{Requesting Feedback Using Solvability Queries}
\label{sec:designtool}

\begin{figure}
    \centering
\includegraphics[width=0.49\textwidth]{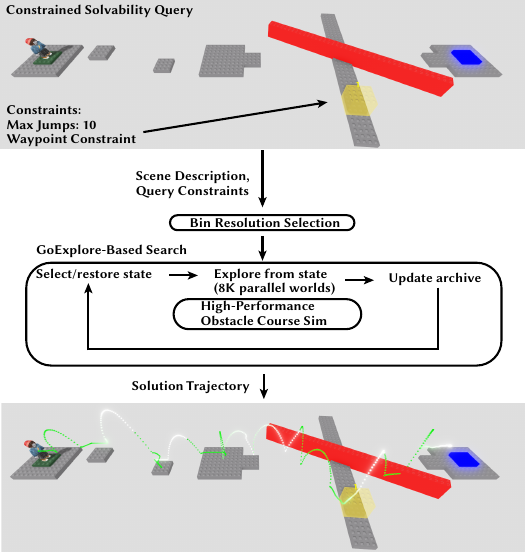}
    \caption{A high-level overview of our system. Queries are submitted from the front-end in Roblox studio and loaded into our GPU-accelerated simulator. We then run parallel GoExplore (using our policy exploration) and return the first solution found. This solutions is visualized as a trajectory in Roblox, and can be replayed synchronously with the motion of the level.}
    \label{fig:algoOVerview}
\end{figure}

Designers use playtesting to gain a better understanding of obstacle \emph{solvability}, or the set of movement sequences a player can or cannot use to solve the obstacle as well as characteristics of those movement sequences (difficulty, strategies, diversity, etc.).
Our system allows designers to query for playthroughs that solve an obstacle while also adhering to designer-specified constraints. Specifically, a \textit{query} specifies a starting region and goal region in an obstacle, as well as a set of constraints that limit the behavior a successful playthrough can exhibit (see Fig.~\ref{fig:algoOVerview}). A successful playthough begins at the centerpoint of the starting region and reaches some point in the goal region while adhering to all constraints. Our system supports the following constraints:

\begin{itemize}[leftmargin=1em]
\item \emph{Waypoint regions}. Regions that successful playthroughs must pass through. A designer uses waypoint region constraints when seeking solutions that take a particular path through the obstacle. An example is shown in Fig.~\ref{fig:algoOVerview}.
\item \emph{Avoid regions}. Regions that successful playthroughs cannot pass through. A designer uses avoid region constraints when they wish to understand how an obstacle can be solved while avoiding a particular path.
\item \emph{Jump count}. Successful playthroughs can only perform a given number of jumps. A designer uses this constraint to support higher level queries about possible play (e.g. what are the \emph{required} jumps along a solution path?). An example is shown in Fig.~\ref{fig:algoOVerview}.

\item \emph{Time limit}. Successful playthroughs must reach the goal region in a specified number of game time steps. A designer can use this constraint to view more efficient playthroughs.
\end{itemize}

Spatial constraints (waypoints and avoid regions) are represented as 3D regions that the designer may adjust using standard translation/rotation/scale primitives. A designer may specify multiple waypoints in sequence, and constrain the number of jumps a player has performed by the time they reach each waypoint.

\paragraph{Visualization of query results}
After submitting a query, the system streams back partial results to the client for immediate visualization.  The system plots both incomplete and successful solutions as 3D trajectories of player movement in the scene.  The designer can interactively select and view trajectories to gain  
insight into \emph{how} a specific playthrough attempt succeeded or \emph{why} it failed. Fig.~\ref{fig:algoOVerview} shows an example query specification 
and one example trajectory that solves the obstacle while adhering to the query's constraints.

%% file: method.tex
\section{Finding Successful Playthroughs}
\label{sec:method}

To be practical for interactive design cycles, our query execution engine must be capable of generating successful playthoughs in seconds, and reliably produce solutions for the diverse and complex set of obstacle segments we expect designers to create. Our solution involves two components: (1) parallel stochastic search over the action space of player movements that optionally is guided by actions sampled from a policy trained offline %
on a large collection of procedurally generated obstacles (see Sec.~\ref{sec:rl-train} and supplement Sec. 7), and (2) a high-performance obstacle game execution engine that executes the exploratory search in parallel on the GPU at over 800K aggregate game steps per second.

The overall flow of the query execution engine is shown in Fig.~\ref{fig:algoOVerview}.  When a query is submitted, query constraints as well as a complete description of the obstacle's 3D scene are communicated to the system. The obstacle is reconstituted in the high-performance obstacle course game simulator, which then executes obstacle playthroughs at high throughput. The simulator terminates playthroughs if they violate query constraints.

\subsection{Policy-Guided Exploratory Search}

One approach to generating solutions to a solvability query is to train a general obstacle-solving agent. We attempted to train such an agent on a large collection of procedurally generated obstacle courses, but found that the resulting policy often failed to solve novel obstacles created by human designers (see results in supplement, Fig. 4). In short, human creativity regularly leads to unexpected diversity and out-of-distribution content. Given the sparse reward structure of fun-to-play obstacles, we also found it impractical to quickly fine-tune a pre-trained policy for a specific query within a latency budget of at most tens of seconds per query (see Sec. 4.3 and Fig. 4 of the supplement). As a result, we chose to base our query execution engine on Go-Explore~\cite{Go-Explore:2021}, a stochastic exploratory search algorithm previously shown to be effective at solving hard exploration problems in simulated environments.

\paragraph{Go-Explore Algorithms}
The Go-Explore family of algorithms depends on two components: a strategy for returning to discovered states, and a strategy for exploring from those states. A learned strategy for returning to discoverd states is Cell-Free Latent Go-Explore~\cite{CFLGE}, which maps discovered states to a latent space learned from a scene dynamics model and returns to states in low-density regions of that latent space. To learn a better exploration strategy, Policy-Based Go-Explore~\cite{Go-Explore:2021} trains a policy to return to previously discovered states that is simultaneously used to explore once returning has been completed. We compare against both these methods in Sec.~\ref{sec:method}.

The main idea of Go-Explore is to perform a randomized search through sequences of actions while maintaining a set of visited states.  In each ``exploration phase'' of the algorithm, Go-Explore selects a prior visited simulation state, restores simulation to this state (the ``go'' part of the algorithm), and resumes exploration for a fixed number of actions from this state (the ``explore'' part). Since it is impractical to store all states visited, Go-Explore discretizes the continuous state space of the obstacle game into a fixed number of bins and stores one game state for each bin. By biasing bin selection towards the least explored bins, Go-Explore rapidly generates action sequences that explore the state space of the game. For our queries, we additionally upweight bins that are closer to the goal region or have reached a waypoint constraint (if a query contains waypoint constraints). Alg.~\ref{alg:Go-Explore} shows the full Go-Explore algorithm. We extend the algorithm with query-dependent binning and a strategy for selective exploration with a pretrained policy, both of which we describe below.

\paragraph{Query-dependent binning} Our system automatically generates binning functions that are specialized to the current obstacle and query. Game states are binned by the XYZ position of the player, the elapsed time, the number of jumps the player has taken, and the position of interactable objects. Because obstacle course games typically require precise lateral movement, we allocate highest resolution to the X and Y dimensions of player position and lower resolution to the Z dimension. We allocate bins to the time, jumps, and interactable object dimensions only if the query features animate moving objects, limited jumps, or interactable objects (see \texttt{binningFunction} in the supplemental code listing).  %

\paragraph{Selecting the exploration policy}. The baseline version of Go-Explore randomly chooses actions during exploration rollouts. This exploration policy may fail to find solutions when obstacles require long sequences of actions to complete, or require solution paths with low margin for error.  

To increase Go-Explore's sample efficiency, prior work trains a policy to guide exploration from saved states.  In the Policy-Based Go-Explore~\cite{Go-Explore:2021} variant of the algorithm, this policy is trained during Go-Explore search by learning to imitate saved action sequences. Intuitively, this approach assumes skills learned to navigate robustly to saved states on a given obstacle will generalize to aid navigation of previously unexplored regions. However, we found that such skills could not be learned quickly enough to accelerate search within our time budget (see Sec.~\ref{sec:query-perf}).

For this reason, we investigate search guidance using a policy that is pretrained on a large collection of procedurally generated obstacles (Sec.~\ref{sec:rl-train}). Our procedurally generated obstacles were designed to model common design patterns and idioms seen in published obstacle course games on popular game platforms (e.g., traversing long walkways, jumping between blocks with air gaps in between, jumping over a wall). During the explore phase of Go-Explore search, we use the policy's actions for the duration of the ``exploration phase" from a restored state if this is the first time that state has been sampled for exploration. Subsequent explorations from the same bin fall back to random actions. Intuitively, this hybrid approach exploits the pretrained policy's skills whenever the search encounters structures that resemble common obstacle course idioms, but dynamically falls back to brute-force random search when the policy's actions fail to produce progress from a state. We describe this method, which we call Go-Explore Policy-Selection, in the following section and evaluate it in Section~\ref{sec:results}.

\subsubsection{Go-Explore with Policy Selection}
\label{sec:ge-ps}
The Go-Explore~\cite{Go-Explore:2021} search policy chooses actions at random with a 95\% probability of repeating the previous action. We call this \emph{default random exploration}.
We modify Go-Explore's explore phase to explore using either our pretrained policy or default random exploration. Specifically, the first time a bin is selected for exploration after being discovered or updated, we use our pretrained policy for the duration of the exploration step. For subsequent selections of this bin, we use default random exploration. This heuristic of using the pretrained policy only once for a new bin is based on two observations: a) our environments are deterministic up to collision resolution in the rigid body simulator, and b) the highest performing policy (at the end of pretraining) outputs nearly deterministic probability distributions over actions even in novel environments. These two observations together suggest that it is likely that the benefit from a pretrained policy can be had with a single rollout from each new state (which in practice might be multiple rollouts if many worlds sample this bin simultaneously). Our policy selection appears on lines 9-14 of Alg.~\ref{alg:Go-Explore}.

\begin{algorithm}[!htbp]
\SetKwFunction{FExplorePhase}{ExplorePhase}
\SetKwFunction{FSolved}{Solved}
\SetKwFunction{FSelectBin}{SelectBin}
\SetKwFunction{FSetSimState}{SetSimulatorState}
\SetKwFunction{FExplore}{Explore}

\SetKwProg{Fn}{Function}{:}{}
\SetKwProg{Pr}{Procedure}{:}{}
\caption{Go-Explore with Policy Selection}
\label{alg:Go-Explore}
\SetAlgoLined
\DontPrintSemicolon

\Pr{\FExplorePhase{$s_{start}$}}{
    \textbf{Input:} Initial high-dimensional simulator state ($s_{start}$). \\
    \textbf{Output:} Archive of all different bins and their best trajectories found ($\mathcal{A}$). \\
    $\mathcal{A} \gets \{ \text{bin}(s_{start}) \mapsto
    (s_{start}, \text{stats}.{init}()) \}$

    \While{not \FSolved{}}{
        $c \gets$ \FSelectBin{$\mathcal{A}$}\;

        $s \gets \mathcal{A}[c].\text{state}$

        \FSetSimState{$s$}\;
        
        $policy$ = \texttt{Random} \\
\tcc{Our Policy Selection}
        \If{$\text{stats}.times\_chosen\_since\_update == 0$}
        {$policy=\texttt{PPO-Policy}$
        }
\tcc{End Our Policy Selection}
        $\text{stats}.times\_chosen\_since\_update \gets \text{stats}.times\_chosen\_since\_update + 1$ \\
        $\text{stats}.times\_chosen \gets \text{stats}.times\_chosen + 1$ \\

        $\mathcal{T}, Solved \gets$ \FExplore{$s, policy$}\;
        
        \ForEach{$(s', r_{cum}, l_{traj}) \in \mathcal{T}$}{
            $c' \gets \text{bin}(s')$

            \eIf{$c' \notin \mathcal{A}$}{
                $\mathcal{A}[c'] \gets
                (s', r_{cum}, l_{traj}, \text{stats}.{init}())$
            }{
                $(s_{old}, r_{old}, l_{old}, \text{stats}) \gets \mathcal{A}[c']$

                \If{$r_{cum} > r_{old} \lor
                (r_{cum} == r_{old} \land l_{traj} < l_{old})$}{
                    $\text{stats}.times\_chosen\_since\_update \gets 0$ \\
                    $\mathcal{A}[c'] \gets
                    (s', r_{cum}, l_{traj}, \text{stats})$
                }
            }
            $\mathcal{A}[c'].\text{stats}.times\_seen \gets \mathcal{A}[c'].\text{stats}.times\_seen + 1$
        }
    }
}
\end{algorithm}

\subsubsection{Exploration Policy Training}\label{sec:rl-train}
We model the task of navigating an obstacle course as a stochastic Markov decision process. Formally, $\mathcal{M} = (\mathcal{S}, \mathcal{A}, \mathcal{P}, \mathcal{R})$, where $\mathcal{S}$ is the state space of the game, $\mathcal{A}$ is the agent's action space, $\mathcal{P}$ is the transition probability function between game states given the agent's action, and $\mathcal{R}$ is the reward function that provides a real-valued score for each action chosen by the agent.
We train our agents using reinforcement learning, which aims to produce a \textit{policy} $\pi : \mathcal{S} \rightarrow \Delta(\mathcal{A})$ that maximizes the expected cumulative reward $J(\pi) = \mathbb{E}\Big[\sum_{t=0}^\infty \gamma^t \mathcal{R}(s_t, a_t) \mid s_0, \pi\Big]$, where $s_t, a_t$ are the state and action at time step $t$ and $\gamma$ is a discount factor on future rewards. We train using PPO\cite{Schulman_PPO} over a collection of procedurally generated obstacle course game environments (see supplement, Sec. 7). Details of the observation, action, and reward space are given in Sec. 3.2 of the supplement.

\subsection{High-Throughput Parallel Search}

To meet the latency requirements of our system, is it critical to execute Go-Explore at exceptionally high throughput. Using a standard game engine platform (in our case, Roblox) to execute game play does not provide sufficient throughput to rapidly find solutions. Instead, we implement a second, high-performance obstacle course game simulator in the Madrona engine~\cite{madrona}.  Madrona is designed for parallel, high-throughput game environment simulation on the GPU. In our tool, when a query is issued, the system serializes the obstacle in Roblox (the geometry of all blocks, along with their motion and behavior) and reconstitutes a close approximation to the game in Madrona. We tune the Madrona physics engine and character controller to approximate the behavior of Roblox so playthroughs executed in the Madrona version of an obstacle yield results that correspond to those observed when executing the same actions in Roblox. Specifically, we calibrate by recording trajectories of agent actions and positions over a simple level in Roblox, then load identical geometry into our simulator and search in parallel over physics parameters to find the combination that best aligns the resulting positions when the action recorded from the Roblox trajectories are replayed. This process is repeated until a certain error threshold across each trajectory is reached.

When a query is submitted, we execute 8K parallel simulations on the GPU, yielding an aggregate effective performance (including the costs of Go-Explore search management) of over 800K game steps per second on an RTX~4090~GPU. \emph{Therefore, in 10 seconds of wall-clock time, our query execution engine can execute and provide feedback based on over 39~hours of simulated obstacle game play.}

%% file: results.tex
\section{Results}
\label{sec:results}

We study how solvability feedback provided by our design tool impacted the process of designing obstacle course games. We first present a case study undertaken by \emph{the authors} that illustrate specific ways in which feedback from our system accelerated and influenced the design process (Sec.~\ref{sec:design-case-studies}). Further studies are included in the supplemental material, with animated walkthroughs presented in our supplemental video. We summarize the results of a user study where four external game designers (three professional, one hobbyist) use the system to create obstacle courses (Sec. ~\ref{sec:user-design-studies}). We discuss one designer's experience in detail and include discussions for all designers in the supplemental material. Finally, we present results that suggest that real human players play the obstacle courses in the way the designers intended (Sec.~\ref{sec:user-gameplay-study}).

\begin{figure}[t]
    \centering
    \includegraphics[width=\linewidth]{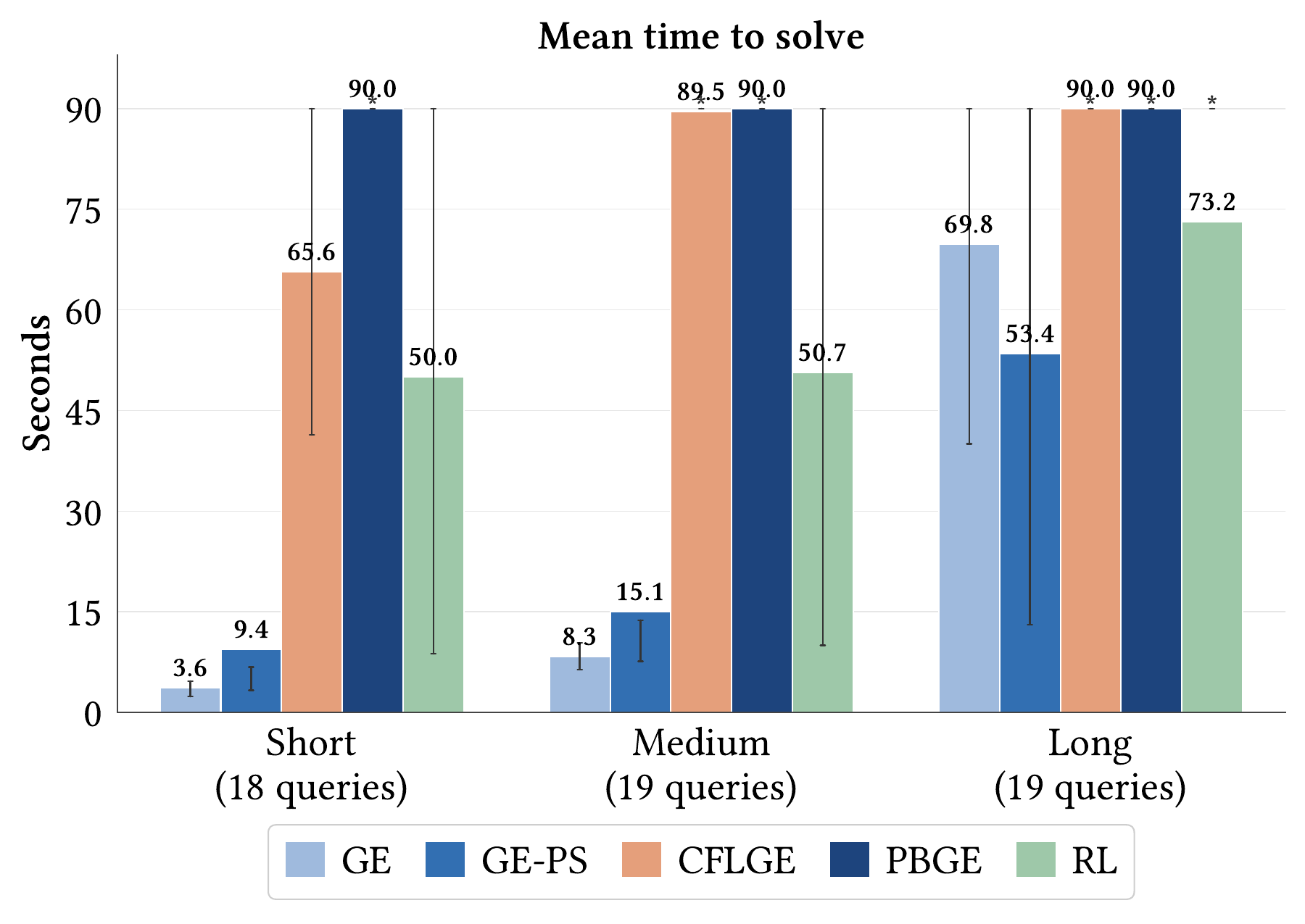}
    \vspace{-2em}
    \caption{We report mean time (with inter-quartile range at 25\%-ile) to solve for 5 methods on queries grouped into tertiles based on the average time to first solution for Go-Explore (Short/Medium/Long). On Long queries, our policy-selection strategy (GE-PS) solves more queries than Go-Explore (GE) leading to a >15s reduction in average solve time at the cost of minimal overhead for Short and Medium queries. Both GE and GE-PS outperform Cell-Fre Latent Go-Explore (CFLGE), Policy-Based Go-Explore (PBGE), and an RL baseline (RL). For details of these ablations, see Sec. 4 of the supplemental material.}
    \label{fig:searchPerformance}
\end{figure}

\subsection{Illustrative Design Case Study}
\label{sec:design-case-studies}

\begin{figure*}[]
    \centering
\includegraphics[width=\textwidth]{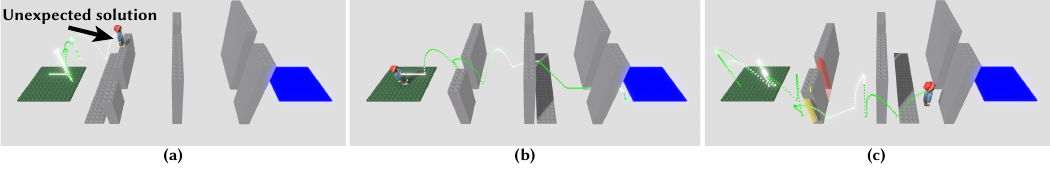}
    \vspace{-2em}
    \caption{Design study 1: Solvability feedback helps the designer identify and remove an undesirable solution. \textbf{(a)} Initial design supports a solution which jumps over the wall instead of passing through its gap. \textbf{(b)} Designer raises the wall so that it cannot be jumped over. \textbf{(c)} Designer verifies an alternate solution (jumping \emph{around} the wall) for experienced players. Please refer to the supplemental video for more details.}
    \label{fig:RidePlatformThroughDoor}
\end{figure*}
In this design study, the designer intends for the player to ride on a thin platform while repositioning to fit through gaps in the walls to avoid being knocked off (Fig. ~\ref{fig:RidePlatformThroughDoor}). The designer wants to verify that this solution is possible and examine alternate solutions to decide which ones to support.

The designer first queries to see if there is a solution and if that solution reflects intended play. The system returns a solution trajectory showing the character riding the platform through the gaps, indicating that intended solution is supported. But the system also shows a different solution (Fig. ~\ref{fig:RidePlatformThroughDoor}, (a), video 1:41). This alternate solution indicates that the first barrier can be jumped over. This runs counter to the designer's intent, so they make the wall higher, and query again to test that their edit prevented the undesirable solution (video 2:13). This query does not return a solution, so the designer concludes that their edit was effective.

The designer now checks for an alternate path that solves the level by jumping around the barrier instead of maneuvering through it, which they construct with avoid region and waypoint primitives. The system returns a trajectory showing that this solution does exist (Fig. ~\ref{fig:RidePlatformThroughDoor}, (c), video 2:57). While this is not the intended solution, it is counterintuitive to a new player \textit{and} mechanically challenging, so the designer does not make any edits to remove it. With these queries, the designer has exhausted the solution classes they care to explore, so this is the final design.

In this example, the system answered designer questions in a way that allowed them to control the solution space by making edits to ensure that an undesirable solution was impossible.

Our further design case studies illustrate the use of our system to explore multiple solution paths using query primitives and evaluating solution properties contributing to difficulty. Multiple additional case studies are included in the supplemental material.

\subsection{Design User Study}
\label{sec:user-design-studies}
We recruited four external designers to use and give feedback on our system. The subjects were: a professional game developer with previous experience in games research (Designer 1), a professional game developer with 19 years of game design experience (Designer 2), an enthusiast game designer specializing in experimental games (Designer 3), and a professional independent game developer who works on 2D platformer games (Designer 4). All designers were given an overview of the tool and its functionality before being asked to design an obstacle. We discuss Designer 1's experience below. Discussions of the other designers are given in the supplemental (Sec. 8.4).

\paragraph{Designer 1}
\begin{figure*}[]
    \centering
    \includegraphics[width=\textwidth]{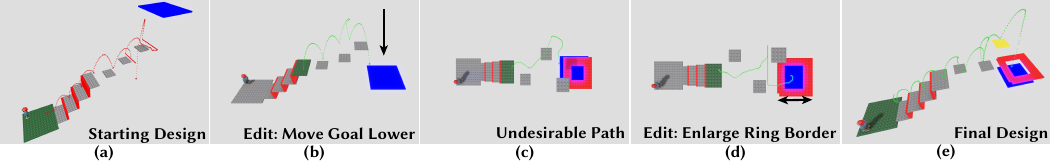}
    \caption{External Designer 1: our system allowed this designer to \textbf{(a)} see why their design was unsolvable, \textbf{(b)} make an edit resulting in a solvable obstacle, \textbf{(c)} see an alternate solution they did not know was possible, \textbf{(d)} make an edit discouraging that solution, and \textbf{(e)} verify that their final design supported their intended solution.}
    \label{fig:designer1}
\end{figure*}

The designer wanted to make something extremely difficult (a "pixel-perfect platformer"). As with the illustrative case study in Sec. ~\ref{sec:design-case-studies}, the designer started by evaluating if their initial design was solvable and what solutions looked like.
Feedback from our system showed that their initial design was unsolvable because the goal platform was too high (Fig. ~\ref{fig:designer1}, a), so the designer made an edit resulting in a solvable design (Fig. ~\ref{fig:designer1}, b).

The designer then added an additional challenge with a ring of dangerous blocks above the goal block. They intended the player to jump from one of the moving blocks through the center of the ring. When they queried to see if this solution was possible, the system revealed an unintended solution that skips the lava ring obstacle entirely (Fig. ~\ref{fig:designer1}, c). This allowed the designer to make edits (shrink goal size, enlarge lava ring radius) that discouraged this unintended solution. Another query showed a solution that did not take the undesirable path (Fig. ~\ref{fig:designer1}, d).

Finally, the designer wished to know if it was possible to jump from the higher platform. To confirm this, they used a waypoint constraint on the top platform to query for a solution trajectory that confirmed their intended gameplay was possible from start to finish (Fig.~\ref{fig:designer1}, e).

Our system allowed the designer to evaluate possible play on obstacles that exceeded what they could comfortably playtest themselves. They commented that seeing the trajectory allowed them to evaluate possible play even without a solution (as in the first query, Fig.~\ref{fig:designer1}, a). They further commented on the utility of finding alternate solutions to their intended one---such a finding caused them to alter their design to be more in line with their goals (Fig.~\ref{fig:designer1}, c to d). Finally, they observed that the interactive feedback was essential for getting the above benefits becuase they could ``press a button and rapidly confirm that a design was still solvable'' after an edit.

\paragraph{External Designer Summary} Like Designer 1, the other external designers used our system to verify desired solutions, discover and prohibit undesired solutions, inspire level edits, and acclimate to a novel 3D creation platform. Further, they were able to make high level inferences about solution trajectories such as their approximate difficulty. Two designers noted that rapid, in-progress trajectory visualizations were crucial for their design process. We discuss studies of the remaining external designers in the supplement.

\subsubsection{Gameplay User Study}
\label{sec:user-gameplay-study}
\begin{figure*}[]
    \centering
\includegraphics[width=\textwidth]{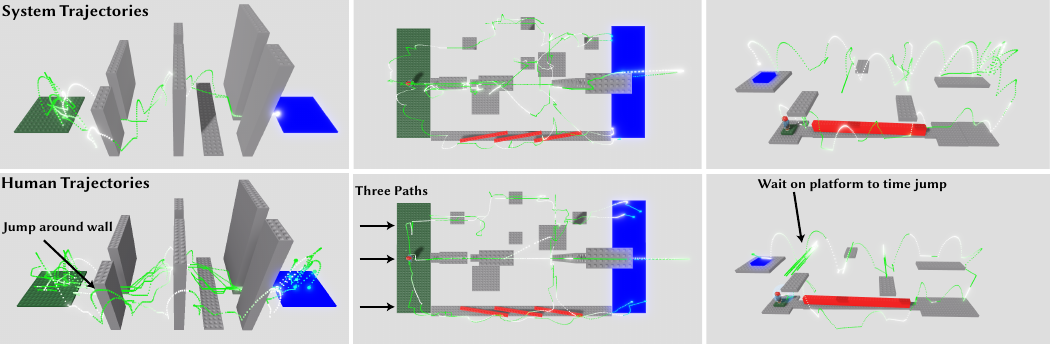}
    \caption{Across three design studies (Sec. ~\ref{sec:design-case-studies}), feedback from our system allowed a designer to create obstacle course designs that supported their desired solutions. These were: a solution that passed through the wall gaps and a solution that jumped around the wall (top, left), a design with multiple solution paths (top, center), and a design where players could time their jumps between platforms to balance the difficulty of the previous section (top, right). All of these solutions were reproduced by human playtesters (bottom row), showing that designers can have confidence that the feedback from our system accurately reflects possible human gameplay. In the far right obstacle, playtesters agreed that the second section was much easier than the first section, as the designer intended based on trajectory feedback. Please refer to the supplemental video to see these obstacle courses in motion.
    }
    \label{fig:human-trajectories}
\end{figure*}

Throughout the design exercises, designers used the tool to gain confidence about how their level design shaped possible play. We show this confidence to be justified with a human playtesting study showing that players do play the game according to the designer's expectations. Our playtesters were college-age students (four male, two female). Though none had extensive experience playing obstacle course games, three had extensive experience with other game genres. In each playtest, the player first attempted the obstacle. If successful, they were prompted to play again, this time after being asked to generate a more difficult solution (e.g., second solution in Fig.~\ref{fig:RidePlatformThroughDoor}) if one existed. Examples for three of our design case studies are shown in Fig.~\ref{fig:human-trajectories}. In each design study, the designer created an obstacle course with a concrete idea of how human play would look on that obstacle. The designer's intent is shown in the trajectory paths in the top row of Fig.~\ref{fig:human-trajectories}. The bottom row of Fig.~\ref{fig:human-trajectories} shows how human playtesters solved the same obstacles. We find that the human players played according to the designer's expectations.

\subsection{Query Performance Evaluation}
\label{sec:query-perf}
A crucial requirement of our system is to provide feedback quickly, so it is important that we identify the most efficient algorithms for generating obstacle solutions. Fig.~\ref{fig:searchPerformance}
measures the mean time-to-first-solution for five techniques: baseline Go-Explore that uses random rollouts (GE)~\cite{Go-Explore:2021}, our modified Go-Explore Policy-Selection algorithm that utilizes guidance from a pretrained policy (GE-PS) (Section~\ref{sec:method}, supplement Sec. 2.4), Policy-based Go-Explore~\cite{Go-Explore:2021} (PBGE), Cell-Free Latent Go-Explore (CFLGE)~\cite{CFLGE}, and solution finding by rolling out the pretrained RL policy with no Go-Explore search (RL).
The figure plots the mean time-to-first-solution for 56 solvability queries logged from our design studies. We split the queries into tertiles based on time taken by GE. Queries that exceed 90~seconds are terminated and treated as completing 90-seconds even though no solution was found.

Our GE baseline is the best performing algorithm for many queries, and often solves queries within a few seconds. For the queries where GE does take longer to solve (``long''), it is advantageous to guide exploration using our pretrained policy. Although GS-PS takes slightly longer on short and medium queries, this is balanced by a significant (>15s on average) speedup for long queries. In effect, our strategy of taking actions using the pretrained policy when restoring from new states recovers much of the limited generalization of the pretrained policy without incurring a high overhead in cases where the policy does not generalize.  Notably, both GE and GE-PS \emph{significantly outperform} variants of the Go-Explore algorithm (PBGE, CFLGE) from recent literature.  These variants involve learning on the fly from experience solving current obstacle, so they incur the cost of policy training but simply do not have time for the learning to have effects on the efficiency of the search. Further, as states further from the start are discovered, PBGE must take additional time to return to discovered states that could be used for exploration. As stated in Section~\ref{sec:method}, the pretrained policy (RL) rarely solved the user-created queries we encountered in our tests. We provide further analysis of additional variants of these algorithms in the supplement, showing GE and GE-PS consistently find solutions faster than all other solution-finding algorithms that we have explored.

%% file: discussion.tex
\section{Discussion}
\label{sec:discussion}
We demonstrate that constrained solution queries allow designers to immediately observe how their design decisions might impact play. In our studies, designers used this rapid feedback to directly modify the obstacle to meet design goals, but it would be interesting to consider other ways playthrough feedback could impact design. For example, we are intrigued by how knowledge of the full-space of potential solution strategies, or the likely failure modes, might help the designer to build player assistance systems or leave breadcrumbs in a design to lead players to a certain solution.

Since players typically traverse obstacles segment by segment, complex obstacle course games are often composed of a series of independent obstacle designs. Once the initial segments are designed, however, the designer must consider global design parameters including obstacle sequencing (e.g. how to create an engaging difficulty progression) and visual communication of function (e.g. make dangerous things appear dangerous, design an obby to suggest a certain solution). It would be fascinating to consider a design tool that accelerated this process, especially automatic evaluation of the solution that is suggested to a player upon first viewing the obstacle.

Finally, we believe that with the increasing ubiquity of AI-generated content, there will soon be an acute need for more tools that help designers turn this abundant 3D content into unique and compelling experiences.  We hope that our work inspires new efforts to explore AI-driven tools that allow creators understand and evaluate how their game design decisions impact game play experiences.

%% file: acknowledgements.tex
\begin{acks}
We thank the participants in our design and gameplay studies for their time and feedback. We thank Nicholas Jennings and Alex Zook for valuable discussions on interactive game design tools. This research is supported by NSF Award \#2219864, the Brown Institute for Media Innovation, and the Stanford Institute for Human-Centered AI (HAI). It is further supported by gifts from Roblox and NVIDIA.
\end{acks}